\documentclass[preprint,aps,12pt,showpacs,nofootinbib,tightenlines]{revtex4}
\usepackage{amsmath}
\usepackage{amssymb}
\usepackage{epsfig}
\usepackage{graphicx}
\begin{document}
\def\pslash{\rlap{\hspace{0.02cm}/}{p}}
\def\eslash{\rlap{\hspace{0.02cm}/}{e}}
\title{The flavor-changing top-charm quark production in the littlest Higgs model with T parity at the ILC}
\author{Xuelei Wang}\email{wangxuelei@sina.com}
\author{Huiling Jin}
\author{Yanju Zhang}
\author{Yanhui Xi}
\affiliation{College of Physics and Information Engineering, Henan
Normal University, Xinxiang, Henan 453007. P.R. China}
\date{\today}
\begin{abstract}
With high luminosity and energy at the ILC and clean SM
backgrounds, the top-charm production at the ILC should have
powerful potential to probe new physics. The littlest Higgs model
with discrete symmetry named "T-parity"(LHT) is one of the most
promising new physics models. In this paper, we study the FC
processes $e^+e^-(\gamma\gamma)\rightarrow t\bar{c}$ at the ILC in
the LHT model. Our study shows that the LHT model can make a
significant contribution to these processes. When the masses of
mirror quarks become large, these two processes are accessible at
the ILC. So the top-charm production at the ILC provides a unique
way to study the properties of the FC couplings in the LHT model
and furthermore test the model.
\end{abstract}


\maketitle


\section{ Introduction}
A simple doublet scalar field yields a perfectly appropriate gauge
symmetry breaking pattern in the Standard Model(SM). On the other
hand, its theoretical shortcomings, such as quadratic
divergencies(hierarchy problem) or the triviality of a $\phi^{4}$
theory suggest that it is embedded in a larger scheme. Recently,
an alternative known as the little Higgs mechanism\cite{little
Higgs}, has been propased. Such mechanism that makes the Higgs
"little" in the current reincarnation of the PGB idea is
collective symmetry breaking. Collective symmetry breaking
protects the Higgs by several symmetries under each of which the
Higgs is an exact Goldstone. Only if the symmetries are broken
collectively, i.e. by more than one coupling in the theory, can
the Higgs pick up a contribution to its mass and hence all
one-loop quadratic divergences to the Higgs mass are avoided. The
most compact implementation of the little Higgs mechanism is known
as the littlest Higgs model\cite{LH}. In this model, the SM is
enlarged to incorporate an approximate $SU(5)$ global symmetry.
This symmetry is broken down to $SO(5)$ spontaneously, though the
mechanism of this breaking is left unspecified. The Higgs is an
approximate Goldstone boson of this breaking. In this model there
are new vector bosons, a heavy top quark and a triplet of heavy
scalars in addition to the SM particles. These new particles can
make significant tree-level contributions to the experimental
observables. So the original LH model suffers strong constraints
from electroweak precision data\cite{constraints}. The most
serious constraints result from the tree-level corrections to
precision electroweak observables due to the exchanges of the
additional heavy gauge bosons, as well as from the small but
non-vanishing vacuum expectation value(VEV) of the additional
weak-triplet scalar field.  To solve this problem, a $Z_2$
discrete symmetry named "T-parity" is introduced\cite{LHT}. The
littlest Higgs model with T parity(LHT), requires the introduction
of "mirror fermions" for each SM fermion doublet. The mirror
fermions are odd under T-parity and can be given large masses and
the SM fields are T-even. T parity explicitly forbids any
tree-level contribution from the heavy gauge bosons to the
observables involving only standard model particles as external
states. It also forbids the interactions that induce the triplet
VEV. As a result, in the LHT model, corrections to precision
electroweak observables are generated at loop-level. This implies
that the constraints are generically weaker than in the tree-level
case, and fine tuning can be avoided\cite{scale}.
   In the LHT model, one of the important ingredients of the
mirror sector is the existence of CKM-like unitary mixing
matrices. These mirror mixing matrices parameterize
flavor-changing(FC) interactions between the SM fermions and the
mirror fermions. Such new FC interactions have a very different
pattern from ones present in the SM and can have significant
contributions to some FC processes. As we know, the SM does not
contain the tree-level FC neutral currents, though it can occur at
higher order through radiative corrections. Because of the loop
suppression, these SM FC effects are hardly to be observed. So
this stimulates a lot of efforts in probing new physics via FC
processes\cite{He,Rare top decay,top-charm,wang,D-LHT,tcv-LHT}.
The impact of the FC interactions in the LHT on FC processes such
as neutral meson mixing and rare $K,B$ meson decays are studied in
Refs.\cite{D-LHT}. The FC couplings between the SM fermions and
the mirror fermions can also induce the loop-level $tcV(V=\gamma,
Z,g)$ couplings. The rare top quark decays $t\rightarrow cV$ in
the LHT model have been studies in Ref.\cite{tcv-LHT} and the
study shows that the decays $t\rightarrow cV$ in the LHT model can
be significantly enhanced relative to those in the SM. The FC
couplings $tcV$ can also make contributions to the top-charm
production. In this paper, we will systematically study the
contribution of the FC couplings in the LHT model to the top-charm
production at the ILC, i.e., $e^+e^-(\gamma\gamma)\rightarrow
t\bar{c}$. The motivations to study the FC top-charm production
are as follows:(1)The LHT model is an ideal idea to solve the
hierarchy problem and does not suffer from severe constraints from
precision electroweak measurewments. (2)Due to the extreme
suppression in the SM, the top-charm production can provide the
clean SM background to probe the quantum effect of the LHT model.
(3)Due to its rather clean environment and high luminosity, the
International Linear Collider(ILC) will be an ideal machine to
probe new physics. In such a collider, in addition to $e^{+}e^{-}$
collision, we can also realize photon-photon collision. The FC
top-charm production at the ILC will be a sensitive probe for
different new physics models and one might distinguish different
new physics models with the precise measurement of these processes
at the ILC.

\indent  This paper is organized as follows. In Sec.II, we briefly
review the LHT model. Sec.III presents the detailed calculation of
the production cross sections of the processes. The numerical
results are shown in Sec.IV. We present discussions and
conclusions in the last section.

\section{ A brief review of the LHT model}

The LH model embeds the electroweak sector of the SM in an
$SU(5)/SO(5)$ non-linear sigma model. It begins with a global
$SU(5)$ symmetry with a locally gauged sub-group $[SU(2)\times
U(1)]^2$. The $SU(5)$ symmetry is spontaneously broken down to
$SO(5)$ via a VEV of order $f$. At the same time, the
$[SU(2)\times U(1)]^2$ gauge symmetry is broken to its diagonal
subgroup $SU(2)_L\times U(1)_Y$ which is identified as the SM
electroweak gauge group. From the $SU(5)/SO(5)$ breaking, there
arise 14 Nambu-Goldstone bosons which are described by the matrix
$\Pi$, given explicitly by
\begin {equation}
\Pi=
\begin{pmatrix}
-\frac{\omega^0}{2}-\frac{\eta}{\sqrt{20}}&-\frac{\omega^+}{\sqrt{2}}
&-i\frac{\pi^+}{\sqrt{2}}&-i\phi^{++}&-i\frac{\phi^+}{\sqrt{2}}\\
-\frac{\omega^-}{\sqrt{2}}&\frac{\omega^0}{2}-\frac{\eta}{\sqrt{20}}
&\frac{v+h+i\pi^0}{2}&-i\frac{\phi^+}{\sqrt{2}}&\frac{-i\phi^0+\phi^P}{\sqrt{2}}\\
i\frac{\pi^-}{\sqrt{2}}&\frac{v+h-i\pi^0}{2}&\sqrt{4/5}\eta&-i\frac{\pi^+}{\sqrt{2}}&
\frac{v+h+i\pi^0}{2}\\
i\phi^{--}&i\frac{\phi^-}{\sqrt{2}}&i\frac{\pi^-}{\sqrt{2}}&
-\frac{\omega^0}{2}-\frac{\eta}{\sqrt{20}}&-\frac{\omega^-}{\sqrt{2}}\\
i\frac{\phi^-}{\sqrt{2}}&\frac{i\phi^0+\phi^P}{\sqrt{2}}&\frac{v+h-i\pi^0}{2}&-\frac{\omega^+}{\sqrt{2}}&
\frac{\omega^0}{2}-\frac{\eta}{\sqrt{20}}
\end{pmatrix}
\end{equation}
Here, $H=(-i\pi^+\sqrt{2},(v+h+i\pi^0)/2)^T$ plays the role of the
SM Higgs doublet, i.e. $h$ is the usual Higgs field, $v=246$ GeV
is the Higgs VEV, and $\pi^{\pm},\pi^0$ are the Goldstone bosons
associated with the spontaneous symmetry breaking $SU(2)_L\times
U(1)_Y\rightarrow U(1)_{em}$. The fields $\eta$ and $\omega$ are
additional Goldstone bosons eaten by heavy gauge bosons when the
$[SU(2)\times U(1)]^2$¡¡gauge group is broken down to
$SU(2)_L\times U(1)_Y$. The field $\Phi$ is a physical scalar
triplet with
\begin {equation}
\Phi=
\begin{pmatrix}
-i\phi^{++}&-i\frac{\phi^+}{\sqrt{2}}\\
-i\frac{\phi^+}{\sqrt{2}}&\frac{-i\phi^0+\phi^P}{\sqrt{2}}
\end{pmatrix}
\end{equation}
Its mass is given by
\begin{eqnarray}
m_{\Phi}=\sqrt{2}m_H\frac{f}{v},
\end{eqnarray}
with $m_H$ being the mass of the SM Higgs scalar.

 In the LHT model, a T-parity discrete symmetry is introduced to make the model consistent
 with the
electroweak precision data. Under the T-parity, the field
$\Phi,\omega,$ and $\eta$ are odd, and the SM Higgs doublet $H$ is
even.

For the gauge group $[SU(2)\times U(1)]^2$, there are eight gauge
bosons, $W^{a\mu}_1, B^{\mu}_1,W^{a\mu}_2, B^{\mu}_2$(a=1,2,3). A
natural way to define the action of T-parity on the gauge fields
is
\begin{eqnarray}
W^a_1\Leftrightarrow W^a_2,~~~~~~B_1\Leftrightarrow B_2.
\end{eqnarray}
An immediate consequence of this definition is that the gauge
couplings of the two $SU(2)\times U(1)$ factors have to be equal.

The gauge boson T-parity eigenstates are given by
\begin{eqnarray}
W^a_L=\frac{W^a_1+W^a_2}{\sqrt{2}},~~~~~~B_L=\frac{B_1+B_2}{\sqrt{2}}~~~~(T-even)\\
W^a_H=\frac{W^a_1-W^a_2}{\sqrt{2}},~~~~~~B_L=\frac{B_1-B_2}{\sqrt{2}}~~~~(T-odd)
\end{eqnarray}

From the first step of symmetry breaking $[SU(2)\times
U(1)]^2\rightarrow SU(2)_L\times U(1)_Y$, the T-odd heavy gauge
bosons acquire masses. The masses of the T-even gauge bosons are
generated only through the second step of symmetry breaking
$SU(2)_L\times U(1)_Y\rightarrow U(1)_{em}$. Finally, the mass
eigenstates are given at order $O(v^2/f^2)$ by

\begin{eqnarray}
W^{\pm}_L=\frac{W^1_L\mp
iW^2_L}{\sqrt{2}},~~~~~~W^{\pm}_H=\frac{W^1_H\mp
iW^2_H}{\sqrt{2}}\\
Z_L=cos\theta_WW^3_L-sin\theta_WB_L,
~~~~~~Z_H=W^3_H+x_H\frac{v^2}{f^2}B_H,\\
A_L=sin\theta_WW^3_L+cos\theta_WB_L,
~~~~~~A_H=-x_H\frac{v^2}{f^2}W^3_H+B_H,
\end{eqnarray}
where $\theta_W$ is the usual weak mixing angle and
\begin{eqnarray}
x_H=\frac{5gg'}{4(5g^2-g'^2)},
\end{eqnarray}
with $g,g'$ being the corresponding coupling constants of
$SU(2)_L$ and $U(1)_Y$. The masses of the T-odd gauge bosons are
given by
\begin{eqnarray}
M_{Z_H}\equiv
M_{W_H}=fg(1-\frac{v^2}{8f^2}),~~~~M_{A_H}=\frac{fg'}{\sqrt{5}}(1-5\frac{v^2}{8f^2}),
\end{eqnarray}
The masses of the T-even gauge bosons are given by
\begin{eqnarray}
M_{W_L}=\frac{gv}{2}(1-\frac{v^2}{12f^2}),
~~~~M_{Z_L}=\frac{gv}{2cos\theta_W}(1-\frac{v^2}{12f^2}),
M_{A_L}=0.
\end{eqnarray}

A consistent and phenomenologically viable implementation of
T-parity in the fermion sector requires the introduction of mirror
fermions. The T-even fermion section consists of the SM quarks,
leptons and an additional heavy quark $T_+$. The T-odd fermion
sector consists of three generations of mirror quarks and leptons
and an additional heavy quark $T_-$. Only the mirror quarks
$(u^i_H,d^i_H)$ are involved in this paper. The mirror fermions
get masses
\begin{eqnarray}
m^u_{H_i}=\sqrt{2}\kappa_if(1-\frac{v^2}{8f^2})\equiv
m_{H_i}(1-\frac{v^2}{8f^2}), \\
\nonumber
m^d_{H_i}=\sqrt{2}\kappa_if\equiv m_{H_i},
\end{eqnarray}
where the Yukawa couplings $\kappa_i$ can in general depend on the
fermion species $i$.

The mirror fermions induce a new flavor structure and there are
four CKM-like unitary mixing matrices in the mirror fermion
sector:
\begin{eqnarray}
V_{H_u},~~V_{H_d},~~V_{H_l},~~V_{H_{\nu}}.
\end{eqnarray}
These mirror mixing matrices are involved in the FC interactions
between the SM fermions and the T-odd mirror fermions which are
mediated by the T-odd heavy gauge bosons or the Goldstone bosons.
$V_{H_u}$ and $V_{H_d}$ satisfy the relation
\begin{eqnarray}
V^{\dag}_{H_u}V_{H_d}=V_{CKM}.
\end{eqnarray}
We parameterize the $V_{H_d}$ with three angles
$\theta^d_{12},\theta^d_{23},\theta^d_{13}$ and three phases
$\delta^d_{12},\delta^d_{23},\delta^d_{13}$

\begin {eqnarray}
V_{H_d}=
\begin{pmatrix}
c^d_{12}c^d_{13}&s^d_{12}s^d_{13}e^{-i\delta^d_{12}}&s^d_{13}e^{-i\delta^d_{13}}\\
-s^d_{12}c^d_{23}e^{i\delta^d_{12}}-c^d_{12}s^d_{23}s^d_{13}e^{i(\delta^d_{13}-\delta^d_{23})}&
c^d_{12}c^d_{23}-s^d_{12}s^d_{23}s^d_{13}e^{i(\delta^d_{13}-\delta^d_{12}-\delta^d_{23})}&
s^d_{23}c^d_{13}e^{-i\delta^d_{23}}\\
s^d_{12}s^d_{23}e^{i(\delta^d_{12}+\delta^d_{23})}-c^d_{12}c^d_{23}s^d_{13}e^{i\delta^d_{13}}&
-c^d_{12}s^d_{23}e^{i\delta^d_{23}}-s^d_{12}c^d_{23}s^d_{13}e^{i(\delta^d_{13}-\delta^d_{12})}&
c^d_{23}c^d_{13}
\end{pmatrix}
\end{eqnarray}
The matrix $V_{H_u}$ is then determined through
$V_{H_u}=V_{H_d}V^{\dag}_{CKM}$. As in the case of the CKM matrix
the angles $\theta^d_{ij}$ can all be made to lie in the first
quadrant with $0\leq
\delta^d_{12},\delta^d_{23},\delta^d_{13}<2\pi$.

\section{The top-charm production in the LHT model}
\subsection{The loop-level FC couplings $tcZ(\gamma)$ in the LHT model}
As we have mentioned above, there are FC interactions between SM
fermions and T-odd mirror fermions which are mediated by the T-odd
heavy gauge bosons($A_H,Z_H,W^{\pm}_H$) or Goldstone
bosons($\eta,\omega^0,\omega^{\pm},$). The relevant Feynman rules
can be found in Ref.\cite{D-LHT}. With these FC couplings, the
loop-level FC couplings $tcZ(\gamma)$  can be induced and the
relevant Feynman diagrams are shown in Fig.1.

As we know, each diagram in Fig.1 actually contains ultraviolet
divergence. Because there is no corresponding tree-level
$tcZ(\gamma)$ couplings to absorb these divergences, the
divergences just cancel each other and the total effective
$tcZ(\gamma)$ couplings are finite as they should be. The
effective one loop-level couplings $tcZ(\gamma)$
 can be directly calculated based on
Fig.1. Their explicit forms, $\Gamma^{\mu}_{tc\gamma}(p_t,p_c)$
and $\Gamma^{\mu}_{tcZ}(p_t,p_c)$, are given in Appendix.

The study has shown that the FC couplings $tcV(V=Z,\gamma,g)$ can
largely enhance the branching ratios of rare top quark decays
$t\rightarrow cV$\cite{tcv-LHT}. On the other hand, the couplings
can also contribute to the top-charm production via the processes
$e^+e^-(\gamma\gamma)\rightarrow t\bar{c}$. We will discuss these
processes in the following.

\subsection{The calculation of the cross sections for the processes $e^+e^-(\gamma\gamma)\rightarrow t\bar{c}$ in the LHT model}
 \hspace{1mm}
In the LHT model, the existence of the FC couplings $tcZ(\gamma)$
can induce the process $e^+e^-\rightarrow t\bar{c}$ at loop-level.
The corresponding Feynman diagram is shown in Fig.2(A).

The production amplitudes are
\begin{eqnarray*}
M_A=M^{\gamma}_A+M^{Z}_A,
\end{eqnarray*}
with
\begin{eqnarray}
 M_A^{\gamma}=-eG(p_1+p_2,0)\bar{u_{t}}(p_{3})\Gamma^{u}_{tc\gamma}(p_{3},p_{4})v_{\bar{c}}(p_{4})\bar{v}_{e^{+}}(p_{2})\gamma_{u}u_{e^{-}}
 (p_{1}),
\end{eqnarray}
\begin{eqnarray}
 M_A^{Z}&=&\frac{g}{\cos\theta_W}G(p_1+p_2,M_Z)\bar{u_{t}}(p_{3})\Gamma^{u}_{tcZ}(p_{3},p_{4})v_{\bar{c}}(p_{4})\bar{v}_{e^{+}}(p_{2})\gamma_{u}\\
 \nonumber
 &&[(-\frac{1}{2}+\sin^{2}\theta_W)P_{L}+(\sin^{2}\theta_W)P_{R}]u_{e^{-}}(p_{1}).
\end{eqnarray}
Where $P_L=\frac{1}{2}(1-\gamma_5)$ and
$P_R=\frac{1}{2}(1+\gamma_5)$ are the left and right chirality
projectors. $p_{1},p_{2}$ are the momenta of the incoming
$e^+,e^-$, and $p_{3},p_{4}$ are the momenta of the outgoing final
states top quark and anti-charm quark, respectively. We also
define $G(p, m)$ as $\frac{1}{p^2-m^2}$.

On the other hand, a unique feature of the ILC is that it can be
transformed to $\gamma\gamma$ collision with the photon beams
generated by using the Compton backscattering of the initial
electron and laser beams. In this case, the energy and luminosity
of the photon beams would be the same order of magnitude of the
original electron beams, and the set of final states at a photon
collider is much richer than that in the $e^+e^-$ mode. So the
realization of the photon collider will open a wider window to
probe new physics. In the LHT model, the top-charm quarks can also
be produced through $\gamma\gamma$ collision, and the relevant
Feynman diagrams are shown in Fig.2(B-C). The invariant production
amplitudes of the process $\gamma\gamma\rightarrow t \bar{c}$
 can be written as:
\begin{eqnarray}
M_B=-\frac{2e}{3}G(p_3-p_1,m_c)\bar{u_{t}}(p_{3})\Gamma^{\mu}_{tc\gamma}(p_3,p_3-p_1)\epsilon_{\mu}(p_1)(\pslash_3-\pslash_1+m_c)
\rlap/\epsilon(p_2)\nu_{\bar{c}}(p_{4}),
 \end{eqnarray}
\begin{eqnarray}
M_C=-\frac{2e}{3}G(p_2-p_4,m_t)\bar{u_{t}}(p_{3})\rlap/\epsilon(p_1)(\pslash_2-\pslash_4+m_t)\Gamma^{\nu}_{tc\gamma}(p_{2}-p_{4},p_4)\epsilon_{\nu}(p_{2})
\nu_{\bar{c}}(p_{4}).
 \end{eqnarray}

With the above amplitudes $M_B,~M_C$, we can directly obtain the
production cross section $\hat{\sigma}(\hat{s})$ for the
subprocess $\gamma\gamma\rightarrow t\bar{c}$ and the total cross
sections at the $e^+e^-$ linear collider can be obtained by
folding $\hat{\sigma}(\hat{s})$ with the photon distribution
function $F(x)$ which is given in Ref.\cite{distribution},

\begin{eqnarray}
\sigma_{tot}(s)=\int^{x_{max}}_{x_{min}}dx_{1}\int^{x_{max}}_{x_{min}
x_{max}/x_1}dx_{2} F(x_{1})F(x_{2})\hat{\sigma}(\hat{s}),
\end{eqnarray}
where $s$ is the c.m. energy squared for $e^+e^-$. The subprocess
occur effectively at $\hat{s}=x_1x_2s$, and $x_i$ are the
fractions of the electron energies carried by the photons. The
explicit form of the photon distribution function $F(x)$ is
\begin{eqnarray}
\displaystyle F(x)=\frac{1}{D(\xi)}\left[1-x+\frac{1}{1-x}
-\frac{4x}{\xi(1-x)}+\frac{4x^2}{\xi^2(1-x)^2}\right],
\end{eqnarray}
with
\begin{eqnarray}
\displaystyle D(\xi)=\left(1-\frac{4}{\xi}-\frac{8}{\xi^2}\right)
\ln(1+\xi)+\frac{1}{2}+\frac{8}{\xi}-\frac{1}{2(1+\xi)^2},
\end{eqnarray}
and
\begin{eqnarray}
\xi=\frac{4E_0\omega_{0}}{m^{2}_{e}}.
\end{eqnarray}
$E_0$ and $\omega_0$ are the incident electron and laser light
energies, and $x=\omega/E_0$. The energy $\omega$ of the scattered
photon depends on its angle $ \theta $ with respect to the
incident electron beam and is given by
\begin{eqnarray}
\omega=\frac{E_{0}(\frac{\xi}{1+\xi})}{1+(\frac{\theta}{\theta_{0}})^{2}}.
\end{eqnarray}
Therefore, at $\theta =0,~\omega=E_{0}\xi/(1+\xi)=\omega_{max}$ is
the maximum energy of the backscattered photon, and
 $x_{max}=\frac{\omega_{max}}{E_{0}}=\frac{\xi}{1+\xi}$.

 To avoid unwanted $e^+e^-$ pair production from the collision between
the incident and back-scattered photons, we should not choose too
large $\omega_0$. The threshold for $e^+e^-$ pair creation is
$\omega_{max}\omega_{0} > m^{2}_{e}$, so we require
$\omega_{max}\omega_{0} \leq m^{2}_{e}$. Solving
$\omega_{max}\omega_{0} = m^{2}_{e}$, we find
\begin{eqnarray}
\xi=2(1+\sqrt{2})=4.8.
\end{eqnarray}
For the choice $\xi=4.8,$ we obtain $x_{max}=0.83$ and
$D(\xi_{max})=1.8.$ The minimum value for $x$ is determined by the
production threshold
\begin{eqnarray}
x_{min}=\frac{\hat{s}_{min}}{x_{max}s},~~~\hat{s}_{min}=(m_t+m_c)^{2}.
\end{eqnarray}

Here we have assumed that both photon beams and electron beams are
unpolarized. We also assume that, the number of the backscattered
photons produced per electron is one.

\section{The numerical results of the processes $e^+e^-(\gamma\gamma)\rightarrow t\bar{c}$ in the LHT model}

To obtain numerical results of the cross sections, we calculate
the amplitudes numerically by using the method of
reference\cite{HZ}, instead of calculating the square of the
production amplitudes analytically. This greatly simplifies our
calculations.

There are several free parameters in the LHT model which are
involved in the production amplitudes. They are the breaking scale
$f$, the masses of the mirror quarks $m_{H_i}(i=1,2,3)$(Here we
have ignored the masses difference between up-type mirror quarks
and down-type mirror quarks), and 6
parameters($\theta^d_{12},~\theta^d_{13},~\theta^d_{23},~\delta^d_{12},~\delta^d_{13},~\delta^d_{23}$)
which are related to the mixing matrix $V_{H_d}$. In
Ref.\cite{D-LHT}, the constraints on the mass spectrum of the
mirror fermions have been investigated from the analysis of
neutral meson mixing in the $K,~B$ and $D$ systems. They found
that a TeV scale GIM suppression is necessary for a generic choice
of $V_{H_d}$. However, there are regions of parameter space where
are only very loose constraints on the mass spectrum of the mirror
fermions. Here we study the processes
$e^+e^-(\gamma\gamma)\rightarrow t\bar{c}$ based on the two
scenarios for the structure of the matrix $V_{H_d}$, as in
Ref.\cite{tcv-LHT}. i.e.,

\hspace{1cm} Case I: $V_{H_d}=1,~~~$$V_{H_u}=V^{\dag}_{CKM}$,

\hspace{1cm} Case II:
$s^d_{23}=1/\sqrt{2},~~s^d_{12}=s^d_{13}=0,~~\delta^d_{12}=\delta^d_{23}=\delta^d_{13}=0.$

In both cases, the constraints on the mass spectrum of the mirror
fermions are very relaxed. For the breaking scale $f$, we take two
typical values: 500 GeV and 1000 GeV.

To get the numerical results of the cross sections, we should also
fix some parameters in the SM as $m_{t}=$174.2 GeV, $m_{c}=$1.25
GeV, $s^{2}_{W}=$0.23, $M_{Z}=$91.87 GeV, $\alpha_e=1/128$, and
$v=246$ GeV\cite{parameters}. For the c.m. energies of the ILC, we
choose $\sqrt{s}=500,~1000$ GeV as examples. On the other hand,
taking account of the detector acceptance, we have taken the basic
cuts on the transverse momentum($p_{T}$) and the
pseudo-rapidity($\eta$) for the final state particles
\begin{eqnarray*}
p_{T}\geq20 GeV,
\hspace{1cm}|\eta|\leq 2.5.
\end{eqnarray*}

The numerical results of the processes
$e^+e^-(\gamma\gamma)\rightarrow t\bar{c}$ are summarized in
Figs.3-5, and the anti-top production is also included in our
calculation. In Figs.3-4, we plot the cross sections of the
processes $e^+e^-(\gamma\gamma)\rightarrow t\bar{c}$ as a function
of $M_{H_3}$ for case I and case II, respectively. In case I, the
mixing in the down type gauge and Goldstone boson interactions are
absent. In this case there are no constraints on the masses of the
mirror quarks at one loop-level from the $K$ and $B$ systems and
the constraints come only from the $D$ system. The constraints on
the mass of the third generation mirror quark are very
weak\cite{D-LHT}. For Case I, we take $m_{H_3}$ to vary in the
range of 500-5000 GeV, and fix $m_{H_1}=m_{H_2}$=300 GeV. We can
see from Fig.3 that both cross sections of the processes
$e^+e^-\rightarrow t\bar{c}$ and $\gamma\gamma\rightarrow
t\bar{c}$ rise very fast with the $m_{H_3}$ increasing. This is
because the couplings between the mirror quarks and the SM quarks
are proportion to the masses of the mirror quarks. The dependences
of c.m. energy $\sqrt{s}$ on the cross sections are different
between the two processes. For the process $e^+e^-\rightarrow
t\bar{c}$, the contributions of the LHT model come from s-channel,
so the large c.m. energy $\sqrt{s}$ depresses the cross section.
However, the LHT model makes t-channel contributions to the
process $\gamma\gamma\rightarrow t\bar{c}$ and the large c.m.
energy can enhance its cross section.
 The masses of the heavy gauge bosons and the mirror quarks,
$M_{V_{H}}$ and $m_{H_i}$, are proportion to $f$, but the scale
$f$ is insensitive to the cross sections of both processes because
the production amplitudes are represented in the form of
$m_{H_i}/M_{V_{H}}$. For Case II, the dependence of the cross
sections on $m_{H_3}$ is presented in Fig.4. In this case, the
constraints from the $K$ and B systems are also very weak.
Compared to Case I, the mixing between the second and third
generations are enhanced with the choice of a bigger mixing angle
$s^d_{23}$. Even with stricter constraints on the masses of the
mirror quarks, the large  masses of the mirror quarks can also
enhance the cross sections significantly. The dependence of the
cross sections on the c.m. energy is similar to that in Case I. In
both Case I and case II, the cross section of
$\gamma\gamma\rightarrow t\bar{c}$ is several orders of magnitude
larger than that of $e^+e^-\rightarrow t\bar{c}$. So the process
$\gamma\gamma\rightarrow t\bar{c}$ benefits from a large cross
section.

In order to provide more information for ILC experiments to probe
the LHT model via the top-charm production, we also give out the
transverse momentum distributions of the top quark in Fig.5. We
fix $\sqrt{s}=500$ GeV, $f=500$ GeV, $m_{H_1}=m_{H_2}=300$ GeV,
$m_{H_3}=2500$ GeV for Case I and fix $\sqrt{s}=500$ GeV, $f=500$
GeV, $m_{H_1}=m_{H_2}=1000$ GeV, $m_{H_3}=1500$ GeV for Case II.
We can see that the $p^{T}_t$ distributions of two processes are
very different. The $p^{T}_t$ distribution of the process
$e^+e^-\rightarrow t\bar{c}$ increases with $p^T_t$ increasing,
but the $p^{T}_t$ distribution of the process
$\gamma\gamma\rightarrow t\bar{c}$ decreases sharply with $p^T_t$.
These two processes can provide complementary information in
different transverse momentum space.

\section{Discussions and conclusions }
\hspace{1mm} Due to the GIM mechanism, the top quark FC
interactions are absent at tree-level and extremely small at
loop-level. So the production rate of $t\bar{c}$ process is very
small in the SM and such process can not be observed. In some new
physics models, there exist new FC interactions which can enhance
the cross section of top-charm production significantly. So the
study of top-charm production would play an important role in
probing new physics models. In the LHT model, there exist the FC
couplings between the SM fermions and mirror fermions which can
make large loop-level contributions to the couplings $tcZ(\gamma)$
and greatly enhance the production rates of the $t\bar{c}$
processes at the ILC. In this paper, we study the processes
$e^+e^-(\gamma\gamma)\rightarrow t\bar{c}$ in the framework of the
LHT model at the ILC. We find that the cross sections of these two
processes vary in a very wide range within the parameter space
limited by the neutral meson mixing in the $K,~B$ and $D$ systems.
With heavy mirror quarks, the cross sections of
$e^+e^-(\gamma\gamma)\rightarrow t\bar{c}$ become very large,
specially for the process $\gamma\gamma\rightarrow t\bar{c}$, but
for relative light mirror quarks the cross sections become very
small. If these processes can be observed and their cross sections
can be measured at the ILC, up-limits on the masses of the mirror
quarks can be obtained. If these processes can not be observed,
relative light mirror quarks are favorable by data of the ILC.
Much more $t\bar{c}$ events can be obtained via photon-photon
collision. So more detail information about the FC couplings in
the LHT model should be obtained via $\gamma\gamma\rightarrow
t\bar{c}$.

\section{Acknowledgments}
\hspace{1mm}

We would thank Junjie Cao for useful discussions and providing the
calculation programs. This work is supported by the National
Natural Science Foundation of China under Grant No.10775039,
10575029 and 10505007.

 \newpage
 \begin{center}
   \Large{Appendix: The explicit expressions of the effective $tcZ(\gamma)$ couplings $\Gamma^{\mu}_{tc\gamma},~\Gamma^{\mu}_{tcZ}$}
\end{center}

The effective $tcZ(\gamma)$ couplings
$\Gamma^{\mu}_{tc\gamma},~\Gamma^{\mu}_{tcZ}$ can be directly
calculated based on Fig.1, and they can be represented  in form of
2-point and 3-point standard functions $B_0,B_1,C_{ij}$. Due to
$m_t>>m_c$, we have safely ignored the terms $m_c/m_t$ in the
calculation. On the other hand, the high order $1/f^2$ terms in
the masses of new gauge bosons and in the Feynman rules are also
ignored. $\Gamma^{\mu}_{tc\gamma}$ and $\Gamma^{\mu}_{tcZ}$ are
depended on the momenta of top quark and charm quark($p_t,p_c$).
Here $p_t$ is outgoing and $p_c$ is incoming. The explicit
expressions of $\Gamma^{\mu}_{tc\gamma},~\Gamma^{\mu}_{tcZ}$ are

\begin{eqnarray*}
\Gamma^{\mu}_{tc\gamma}(p_t,p_c)&=&\Gamma^{\mu}_{tc\gamma}(\eta^{0})+\Gamma^{\mu}_{tc\gamma}(\omega^{0})
+\Gamma^{\mu}_{tc\gamma}(\omega^{\pm})+\Gamma^{\mu}_{tc\gamma}(A_{H})+\Gamma^{\mu}_{tc\gamma}(Z_{H})
+\Gamma^{\mu}_{tc\gamma}(W_{H}^{\pm})\\&&+\Gamma^{\mu}_{tc\gamma}(W_{H}^{\pm}\omega^{\pm}),
\end{eqnarray*}
\begin{eqnarray*}
\Gamma^{\mu}_{tc\gamma}(\eta^{0})&=&\frac{i}{16\pi^{2}}\frac{eg^{\prime2}}{150M_{A_{H}}^{2}}
(V_{Hu})_{3i}(V_{Hu})_{i2}{m_{Hi}^{2}}\\&&
\{[B_{0}(-p_{t},m_{Hi},0)-B_{0}(-p_{c},m_{Hi},0)
+B_{1}(-p_{t},m_{Hi},0)\\&&+2C_{24}^{a} -2p_{t}\cdot
p_{c}(C_{12}^{a}+C_{23}^{a})+m_{t}^{2}(C_{21}^{a}+C_{11}^{a}+C_{0}^{a})-m_{Hi}^{2}C_{0}^{a}]
\gamma^{\mu}P_{L}\\&&
+[-2m_{t}(C_{21}^{a}+2C_{11}^{a}+C_{0}^{a})]p_{t}^{\mu}P_{L}+[2m_{t}(C_{23}^{a}+2C_{12}^{a})]p_{c}^{\mu}P_{L}\},
\end{eqnarray*}
\begin{eqnarray*}
\Gamma^{\mu}_{tc\gamma}(\omega^{0})&=&\frac{i}{16\pi^{2}}\frac{eg^{2}}{6M_{Z_{H}}^{2}}
(V_{Hu})_{3i}(V_{Hu})_{i2}{m_{Hi}^{2}}\\&&
\{[B_{0}(-p_{t},m_{Hi},0)-B_{0}(-p_{c},m_{Hi},0)
+B_{1}(-p_{t},m_{Hi},0)\\&&+2C_{24}^{b} -2p_{t}\cdot
p_{c}(C_{12}^{b}+C_{23}^{b})+m_{t}^{2}(C_{21}^{b}+C_{11}^{b}+C_{0}^{b})-m_{Hi}^{2}C_{0}^{b}]
\gamma^{\mu}P_{L}\\&&
+[-2m_{t}(C_{21}^{b}+2C_{11}^{b}+C_{0}^{b})]p_{t}^{\mu}P_{L}+[2m_{t}(C_{23}^{b}+2C_{12}^{b})]p_{c}^{\mu}P_{L}\},
\end{eqnarray*}
\begin{eqnarray*}
\Gamma^{\mu}_{tc\gamma}(\omega^{\pm})&=&\frac{i}{16\pi^{2}}\frac{eg^{2}}{6M_{W_{H}}^{2}}
(V_{Hu})_{3i}(V_{Hu})_{i2}{m_{Hi}^{2}}\\&&
\{2[(B_{0}(-p_{t},m_{Hi},0)-B_{0}(-p_{c},m_{Hi},0)
+B_{1}(-p_{t},m_{Hi},0))\\&&-2C_{24}^{c}+6C_{24}^{g}+2p_{t}\cdot
p_{c} (C_{12}^{c}+C_{23}^{c})-m_{t}^{2}(C_{21}^{c}
+C_{11}^{c}+C_{0}^{c})+m_{Hi}^{2}C_{0}^{c}] \gamma^{\mu}P_{L}\\&&
+[2m_{t}(C_{21}^{c}+2C_{11}^{c}+C_{0}^{c})+3m_{t}(2C_{21}^{g}+C_{11}^{g})]p_{t}^{\mu}P_{L}\\&&
+[-2m_{t}(C_{23}^{c}+2C_{12}^{c})-3m_{t}(2C_{23}^{g}+C_{11}^{g})]p_{c}^{\mu}P_{L}\},
\end{eqnarray*}
\begin{eqnarray*}
\Gamma^{\mu}_{tc\gamma}(A_{H})&=&\frac{i}{16\pi^{2}}\frac{eg^{\prime2}}{75}
(V_{Hu})_{3i}(V_{Hu})_{i2}\\&&
\{[B_{1}(-p_{t},m_{Hi},M_{A_{H}})+2C_{24}^{d} -2p_{t}\cdot
p_{c}(C_{11}^{d}+C_{23}^{d})+m_{t}^{2}(C_{21}^{d}+C_{11}^{d})\\&&-m_{Hi}^{2}C_{0}^{d}]
\gamma^{\mu}P_{L}
+[-2m_{t}(C_{21}^{d}+C_{11}^{d})]p_{t}^{\mu}P_{L}
+[2m_{t}(C_{23}^{d}+C_{11}^{d})]p_{c}^{\mu}P_{L}\},
\end{eqnarray*}
\begin{eqnarray*}
\Gamma^{\mu}_{tc\gamma}(Z_{H})&=&\frac{i}{16\pi^{2}}\frac{eg^{2}}{3}
(V_{Hu})_{3i}(V_{Hu})_{i2}\\&&
\{[B_{1}(-p_{t},m_{Hi},M_{Z_{H}})+2C_{24}^{e} -2p_{t}\cdot
p_{c}(C_{11}^{e}+C_{23}^{e})+m_{t}^{2}(C_{21}^{e}+C_{11}^{e})\\&&-m_{Hi}^{2}C_{0}^{e}]
\gamma^{\mu}P_{L}
+[-2m_{t}(C_{21}^{e}+C_{11}^{e})]p_{t}^{\mu}P_{L}
+[2m_{t}(C_{23}^{e}+C_{11}^{e})]p_{c}^{\mu}P_{L}\},
\end{eqnarray*}
\begin{eqnarray*}
\Gamma^{\mu}_{tc\gamma}(W_{H}^{\pm})&=&\frac{i}{16\pi^{2}}\frac{eg^{2}}{6}
(V_{Hu})_{3i}(V_{Hu})_{i2}\\&&
\{[4B_{1}(-p_{t},m_{Hi},M_{W_{H}})+2B_{0}(p_{c},m_{Hi},M_{W_{H}})-4C_{24}^{f}
+4C_{24}^{h}\\&&+4p_{t}\cdot
p_{c}(C_{11}^{f}+C_{23}^{f})-2m_{t}^{2}(C_{21}^{f}+C_{11}^{f})+2m_{Hi}^{2}C_{0}^{f}
+2M_{W_{H}}^{2}C_{0}^{h}\\&&-4p_{t}\cdot
p_{c}(C_{11}^{h}+C_{0}^{h})+m_{t}^{2}(3C_{11}^{h}+C_{0}^{h})]
\gamma^{\mu}P_{L}\\&&
+[4m_{t}(C_{21}^{f}+C_{11}^{f})+2m_{t}(3C_{11}^{h}+2C_{21}^{h}+C_{0}^{h})]p_{t}^{\mu}P_{L}\\&&
+[-4m_{t}(C_{23}^{f}+C_{11}^{f})-2m_{t}(2C_{23}^{h}+3C_{12}^{h}-C_{11}^{h}-C_{0}^{h})]p_{c}^{\mu}P_{L}\},
\end{eqnarray*}
\begin{eqnarray*}
\Gamma^{\mu}_{tc\gamma}(W_{H}^{\pm}\omega^{\pm})&=&\frac{i}{16\pi^{2}}\frac{eg^{2}}0{2}
(V_{Hu})_{3i}(V_{Hu})_{i2}\\&&\{[m_{Hi}^{2}(C_{0}^{i}-C_{0}^{j})+m_{t}^{2}(C_{11}^{j}+C_{0}^{j})]
\gamma^{\mu}P_{L}+[-2m_{t}C_{12}^{j}]p_{c}^{\mu}P_{L}\}.
\end{eqnarray*}
\begin{eqnarray*}
\Gamma^{\mu}_{tcZ}(p_t,p_c)&=&\Gamma^{\mu}_{tcZ}(\eta^{0})+\Gamma^{\mu}_{tcZ}(\omega^{0})
+\Gamma^{\mu}_{tcZ}(\omega^{\pm})+\Gamma^{\mu}_{tcZ}(A_{H})+\Gamma^{\mu}_{tcZ}(Z_{H})
+\Gamma^{\mu}_{tcZ}(W_{H}^{\pm})\\&&+\Gamma^{\mu}_{tcZ}(W_{H}^{\pm}\omega^{\pm}),
\end{eqnarray*}
\begin{eqnarray*}
\Gamma^{\mu}_{tcZ}(\eta^{0})&=&\frac{i}{16\pi^{2}}\frac{g}{\cos\theta_{W}}
(\frac{1}{2}-\frac{2}{3}\sin^{2}\theta_{W})\frac{g^{\prime2}}{100M_{A_{H}}^{2}}
(V_{Hu})_{3i}(V_{Hu})_{i2}{m_{Hi}^{2}}\\&&
\{[B_{0}(-p_{t},m_{Hi},0)-B_{0}(-p_{c},m_{Hi},0)
+B_{1}(-p_{t},m_{Hi},0)\\&&+2C_{24}^{a} -2p_{t}\cdot
p_{c}(C_{12}^{a}+C_{23}^{a})+m_{t}^{2}(C_{21}^{a}+C_{11}^{a}+C_{0}^{a})-m_{Hi}^{2}C_{0}^{a}]
\gamma^{\mu}P_{L}\\&&
+[-2m_{t}(C_{21}^{a}+2C_{11}^{a}+C_{0}^{a})]p_{t}^{\mu}P_{L}+[2m_{t}(C_{23}^{a}+2C_{12}^{a})]p_{c}^{\mu}P_{L}\},
\end{eqnarray*}
\begin{eqnarray*}
\Gamma^{\mu}_{tcZ}(\omega^{0})&=&\frac{i}{16\pi^{2}}\frac{g}{\cos\theta_{W}}
(\frac{1}{2}-\frac{2}{3}\sin^{2}\theta_{W})\frac{g^{2}}{4M_{Z_{H}}^{2}}
(V_{Hu})_{3i}(V_{Hu})_{i2}{m_{Hi}^{2}}\\&&
\{[B_{0}(-p_{t},m_{Hi},0)-B_{0}(-p_{c},m_{Hi},0)
+B_{1}(-p_{t},m_{Hi},0)\\&&+2C_{24}^{b} -2p_{t}\cdot
p_{c}(C_{12}^{b}+C_{23}^{b})+m_{t}^{2}(C_{21}^{b}+C_{11}^{b}+C_{0}^{b})-m_{Hi}^{2}C_{0}^{b}]
\gamma^{\mu}P_{L}\\&&
+[-2m_{t}(C_{21}^{b}+2C_{11}^{b}+C_{0}^{b})]p_{t}^{\mu}P_{L}+[2m_{t}(C_{23}^{b}+2C_{12}^{b})]p_{c}^{\mu}P_{L}\},
\end{eqnarray*}
\begin{eqnarray*}
\Gamma^{\mu}_{tcZ}(\omega^{\pm})&=&\frac{i}{16\pi^{2}}\frac{g}{\cos\theta_{W}}
\frac{g^{2}}{2M_{W_{H}}^{2}}(V_{Hu})_{3i}(V_{Hu})_{i2}{m_{Hi}^{2}}\\&&
\{[(\frac{1}{2}-\frac{2}{3}\sin^{2}\theta_{W})
(B_{0}(-p_{t},m_{Hi},0)
-B_{0}(-p_{c},m_{Hi},0)\\&&+B_{1}(-p_{t},m_{Hi},0))
+(-\frac{1}{2}+\frac{1}{3}\sin^{2}\theta_{W})(2C_{24}^{c}-2p_{t}\cdot
p_{c}
(C_{12}^{c}+C_{23}^{c})\\&&+m_{t}^{2}(C_{21}^{c}+C_{11}^{c}+C_{0}^{c})-m_{Hi}^{2}C_{0}^{c})
+2\cos^{2}\theta_{W}C_{24}^{g}] \gamma^{\mu}P_{L}\\&&
+[(-\frac{1}{2}+\frac{1}{3}\sin^{2}\theta_{W})
(-2m_{t}(C_{21}^{c}+2C_{11}^{c}+C_{0}^{c}))\\&&+\cos^{2}\theta_{W}m_{t}(2C_{21}^{g}+C_{11}^{g})]p_{t}^{\mu}P_{L}
+[2(-\frac{1}{2}+\frac{1}{3}\sin^{2}\theta_{W})
m_{t}(C_{23}^{c}+2C_{12}^{c})\\&&-\cos^{2}\theta_{W}m_{t}(2C_{23}^{g}+C_{11}^{g})]p_{c}^{\mu}P_{L}\},
\end{eqnarray*}
\begin{eqnarray*}
\Gamma^{\mu}_{tcZ}(A_{H})&=&\frac{i}{16\pi^{2}}\frac{g}{\cos\theta_{W}}
(\frac{1}{2}-\frac{2}{3}\sin^{2}\theta_{W})\frac{g^{\prime2}}{50}
(V_{Hu})_{3i}(V_{Hu})_{i2}\\&&
\{[B_{1}(-p_{t},m_{Hi},M_{A_{H}})+2C_{24}^{d} -2p_{t}\cdot
p_{c}(C_{11}^{d}+C_{23}^{d})+m_{t}^{2}(C_{21}^{d}+C_{11}^{d})\\&&-m_{Hi}^{2}C_{0}^{d}]
\gamma^{\mu}P_{L}
+[-2m_{t}(C_{21}^{d}+C_{11}^{d})]p_{t}^{\mu}P_{L}
+[2m_{t}(C_{23}^{d}+C_{11}^{d})]p_{c}^{\mu}P_{L}\},
\end{eqnarray*}
\begin{eqnarray*}
\Gamma^{\mu}_{tcZ}(Z_{H})&=&\frac{i}{16\pi^{2}}\frac{g}{\cos\theta_{W}}
(\frac{1}{2}-\frac{2}{3}\sin^{2}\theta_{W})\frac{g^{2}}{2}
(V_{Hu})_{3i}(V_{Hu})_{i2}\\&&
\{[B_{1}(-p_{t},m_{Hi},M_{Z_{H}})+2C_{24}^{e} -2p_{t}\cdot
p_{c}(C_{11}^{e}+C_{23}^{e})+m_{t}^{2}(C_{21}^{e}+C_{11}^{e})\\&&-m_{Hi}^{2}C_{0}^{e}]
\gamma^{\mu}P_{L}
+[-2m_{t}(C_{21}^{e}+C_{11}^{e})]p_{t}^{\mu}P_{L}
+[2m_{t}(C_{23}^{e}+C_{11}^{e})]p_{c}^{\mu}P_{L}\},
\end{eqnarray*}
\begin{eqnarray*}
\Gamma^{\mu}_{tcZ}(W_{H}^{\pm})&=&\frac{i}{16\pi^{2}}\frac{g}{\cos\theta_{W}}{g^{2}}(V_{Hu})_{3i}(V_{Hu})_{i2}\\&&
\{[(\frac{1}{2}-\frac{2}{3}\sin^{2}\theta_{W})
B_{1}(-p_{t},m_{Hi},M_{W_{H}})\\&&+(-\frac{1}{2}
+\frac{1}{3}\sin^{2}\theta_{W})(2C_{24}^{f}-2p_{t}\cdot
p_{c}(C_{11}^{f}+C_{23}^{f})+m_{t}^{2}(C_{21}^{f}
+C_{11}^{f})-m_{Hi}^{2}C_{0}^{f})\\&&+\frac{1}{6}\cos^{2}\theta_{W}(2B_{0}(p_{c},m_{Hi},M_{W_{H}})
+4C_{24}^{h}-4p_{t}\cdot
p_{c}(C_{11}^{h}+C_{0}^{h})\\&&+m_{t}^{2}(3C_{11}^{h}+C_{0}^{h})
+2M_{W_{H}}^{2}C_{0}^{h})]\gamma^{\mu}P_{L}\\&&+[(-\frac{1}{2}
+\frac{1}{3}\sin^{2}\theta_{W})(-2m_{t}(C_{21}^{f}+C_{11}^{f}))\\&&
+\frac{1}{3}\cos^{2}\theta_{W}m_{t}(2C_{21}^{h}+3C_{11}^{h}+C_{0}^{h})]p_{t}^{\mu}P_{L}\\&&
+[2(-\frac{1}{2}+\frac{1}{3}\sin^{2}\theta_{W})m_{t}(C_{23}^{f}+C_{11}^{f})\\&&
-\frac{1}{3}\cos^{2}\theta_{W}m_{t}(2C_{23}^{h}+3C_{12}^{h}-C_{11}^{h}-C_{0}^{h})]p_{c}^{\mu}P_{L}\},
\end{eqnarray*}
\begin{eqnarray*}
\Gamma^{\mu}_{tcZ}(W_{H}^{\pm}\omega^{\pm})&=&\frac{i}{16\pi^{2}}g\cos\theta_{W}\frac{g^{2}}{2}
(V_{Hu})_{3i}(V_{Hu})_{i2}\\&&\{[m_{Hi}^{2}(C_{0}^{i}-C_{0}^{j})+m_{t}^{2}(C_{11}^{j}+C_{0}^{j})]
\gamma^{\mu}P_{L}+[-2m_{t}C_{12}^{j}]p_{c}^{\mu}P_{L}\}\\
\end{eqnarray*}
The three-point standard functions $C_0,~C_{ij}$ are defined as
\begin{eqnarray*}
C_{ij}^{a}&=&C_{ij}^{a}(-p_{t},p_{c},m_{Hi},0,m_{Hi}),\\
C_{ij}^{b}&=&C_{ij}^{b}(-p_{t},p_{c},m_{Hi},0,m_{Hi}),\\
C_{ij}^{c}&=&C_{ij}^{c}(-p_{t},p_{c},m_{Hi},0,m_{Hi}),\\
C_{ij}^{d}&=&C_{ij}^{d}(-p_{t},p_{c},m_{Hi},M_{A_{H}},m_{Hi}),\\
C_{ij}^{e}&=&C_{ij}^{e}(-p_{t},p_{c},m_{Hi},M_{Z_{H}},m_{Hi}),\\
C_{ij}^{f}&=&C_{ij}^{f}(-p_{t},p_{c},m_{Hi},M_{W_{H}},m_{Hi}),\\
C_{ij}^{g}&=&C_{ij}^{g}(-p_{t},p_{c},0,m_{Hi},0),\\
C_{ij}^{h}&=&C_{ij}^{h}(-p_{t},p_{c},M_{W_{H}},m_{Hi},M_{W_{H}}),\\
C_{ij}^{i}&=&C_{ij}^{i}(-p_{t},p_{c},M_{W_{H}},m_{Hi},0),\\
C_{ij}^{j}&=&C_{ij}^{j}(-p_{t},p_{c},0,m_{Hi},M_{W_{H}}).
\end{eqnarray*}

\newpage
\begin{figure}
\begin{center}
\includegraphics [scale=0.7] {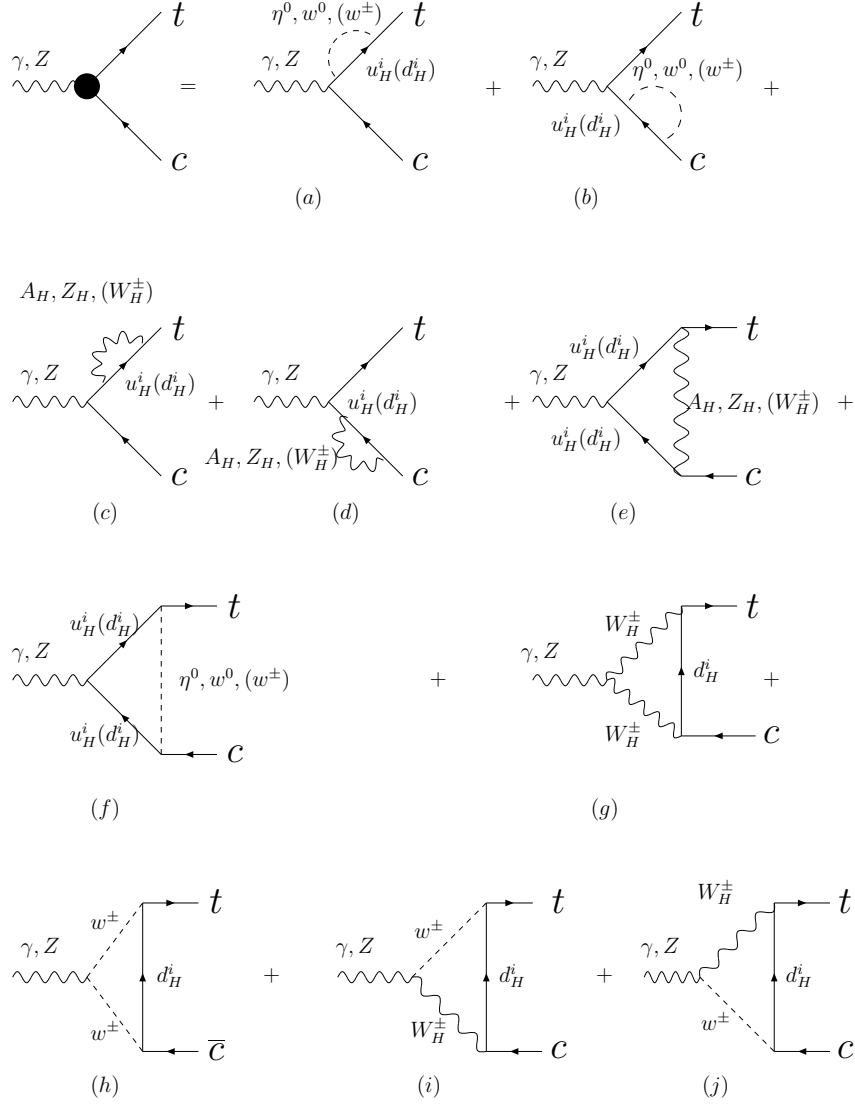}
\caption{The Feynman diagrams of the one-loop contributions of the
LHC model to the couplings $tcZ(\gamma)$.} \label{fig:fig1}
\end{center}
\end{figure}

\newpage
\begin{figure}
\begin{center}
\includegraphics [scale=0.7] {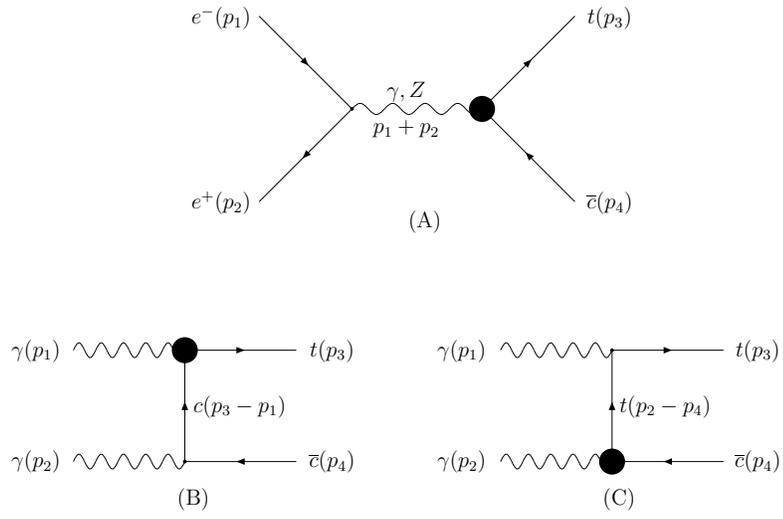}
\caption{The Feynman diagrams of the processes
$e^+e^-(\gamma\gamma)\rightarrow t\bar{c}$ in the LHT model.}
\label{fig:fig2}
\end{center}
\end{figure}

\newpage
\begin{figure}[h]
\scalebox{0.7}{\epsfig{file=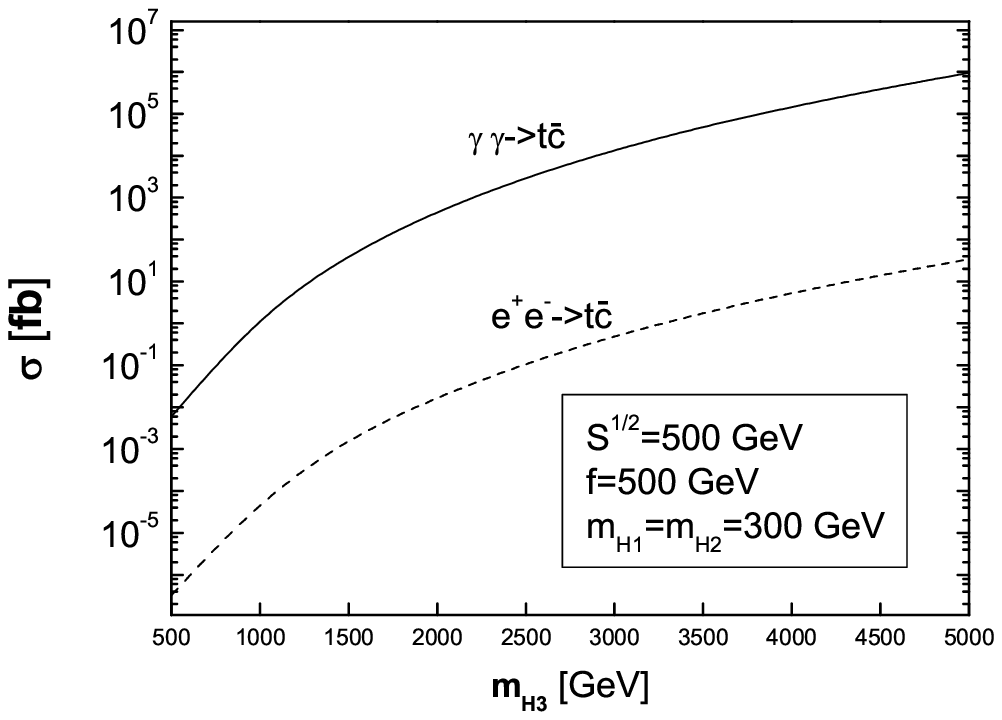}}
\scalebox{0.7}{\epsfig{file=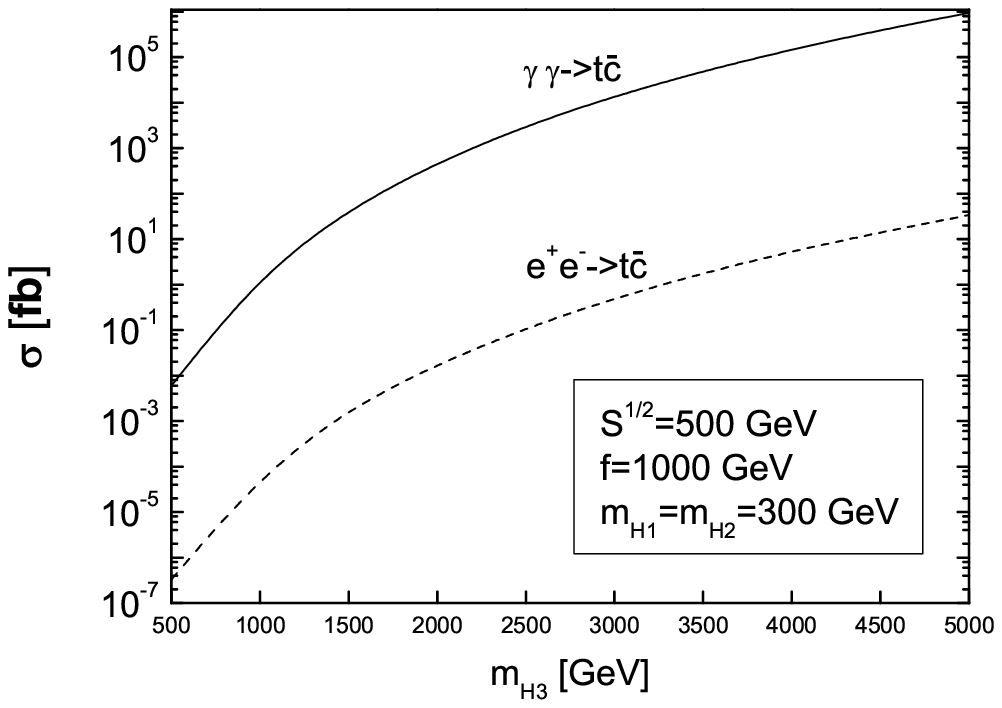}}\\
\scalebox{0.7}{\epsfig{file=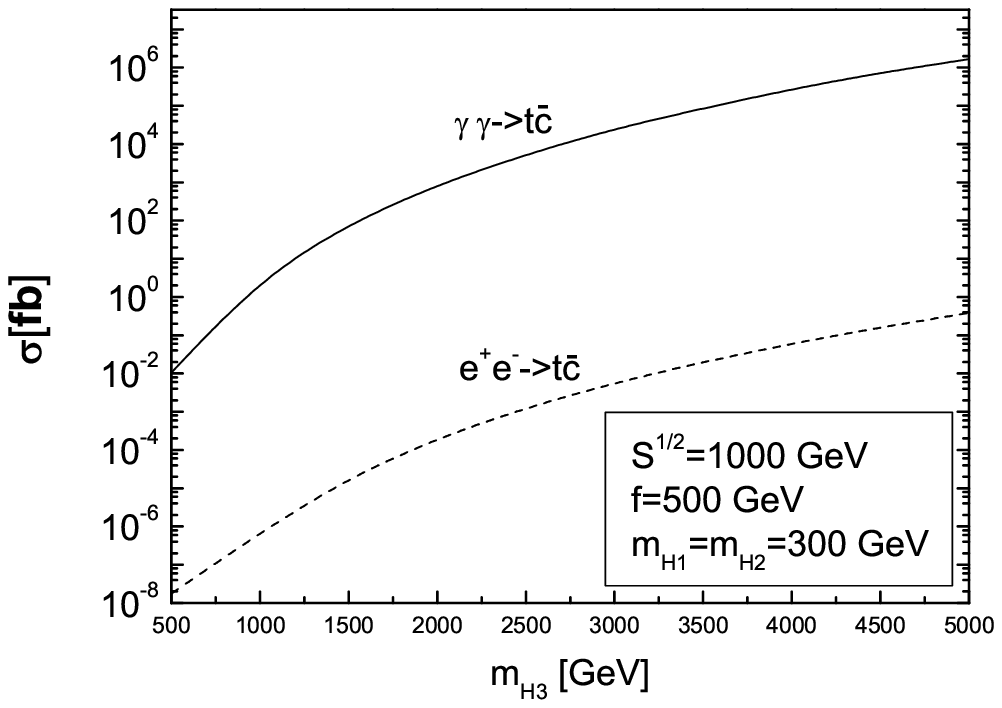}}
\scalebox{0.7}{\epsfig{file=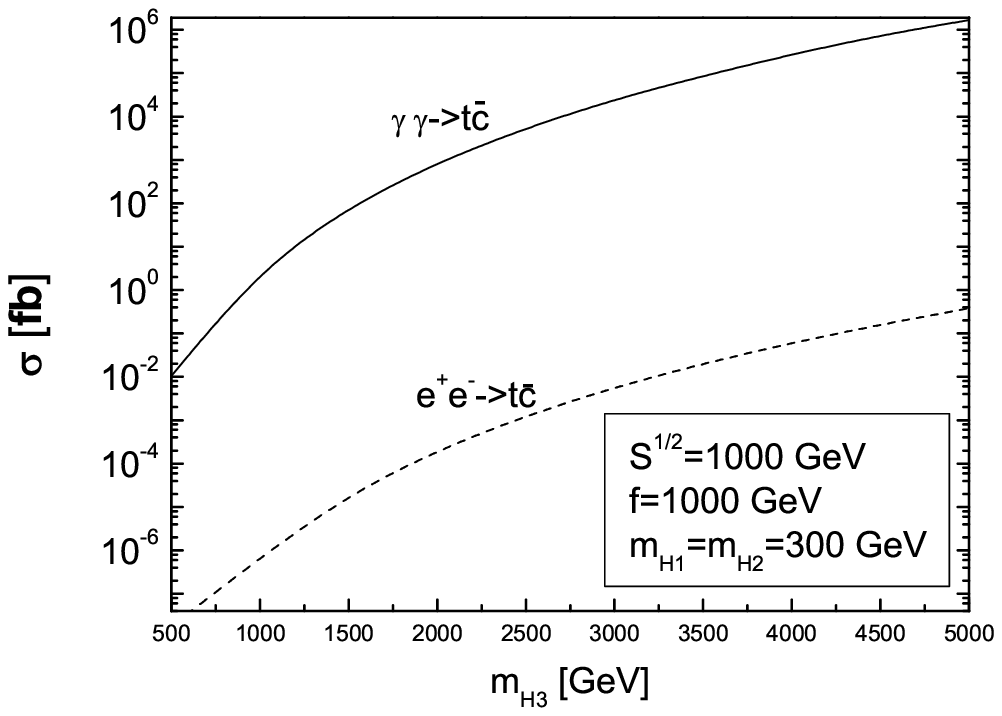}}\\
\caption{\small The cross sections of the processes
$e^+e^-(\gamma\gamma)\rightarrow t\bar{c}$ in the LHT model for
Case I, as a function of $M_{H_3}$.}
\end{figure}

\newpage

\begin{figure}[h]
\scalebox{0.7}{\epsfig{file=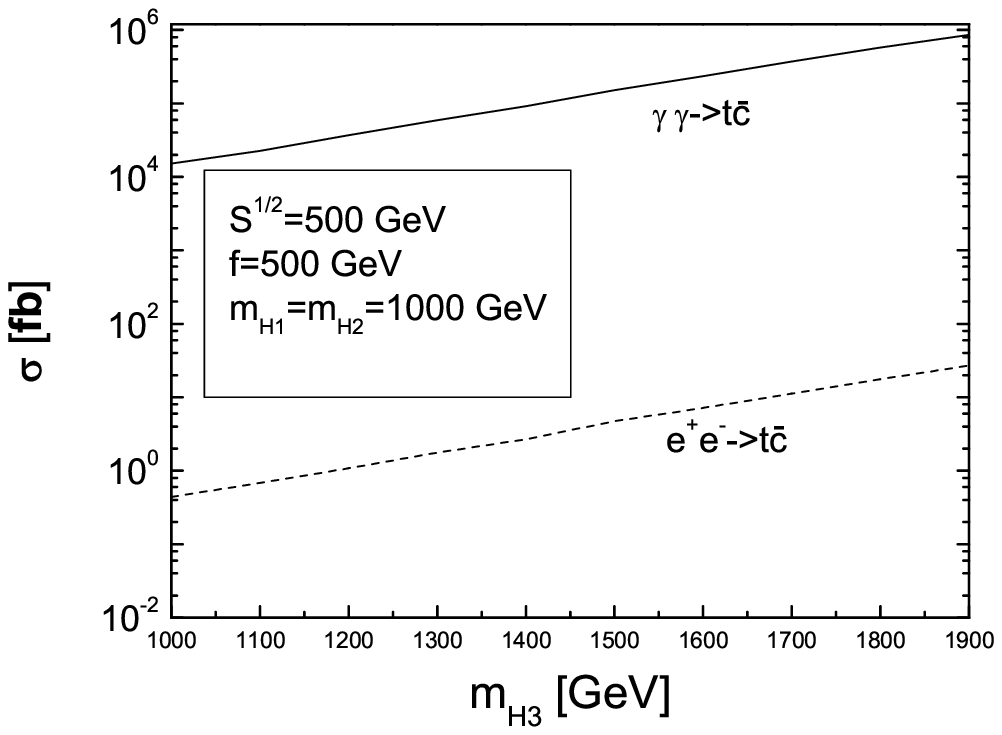}}
\scalebox{0.7}{\epsfig{file=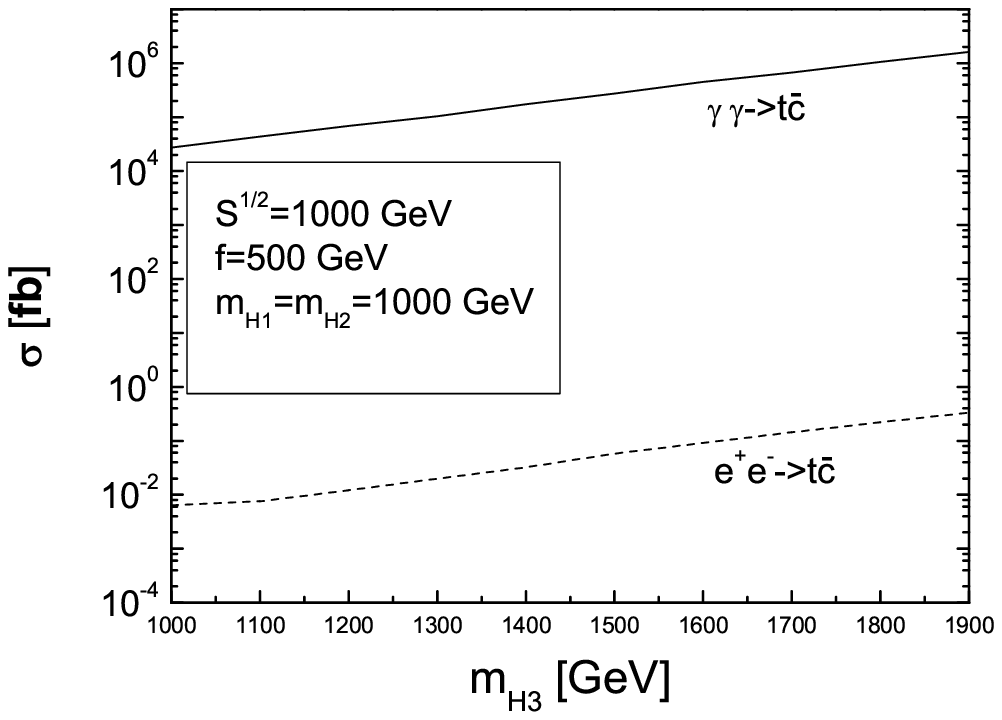}}\\
\scalebox{0.7}{\epsfig{file=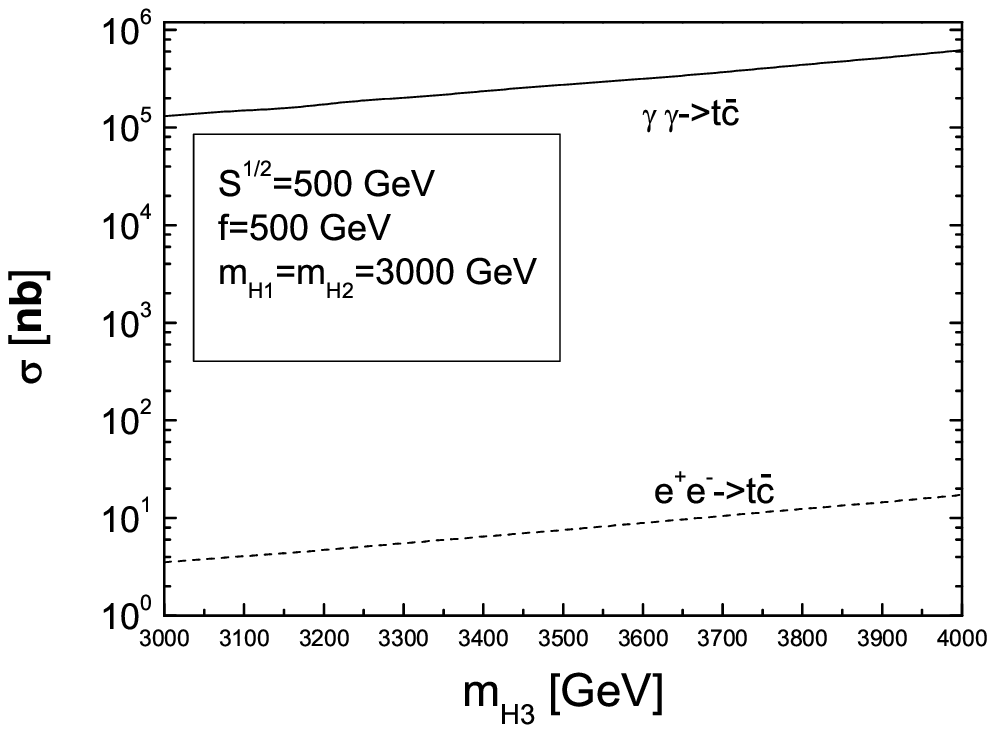}}
\scalebox{0.7}{\epsfig{file=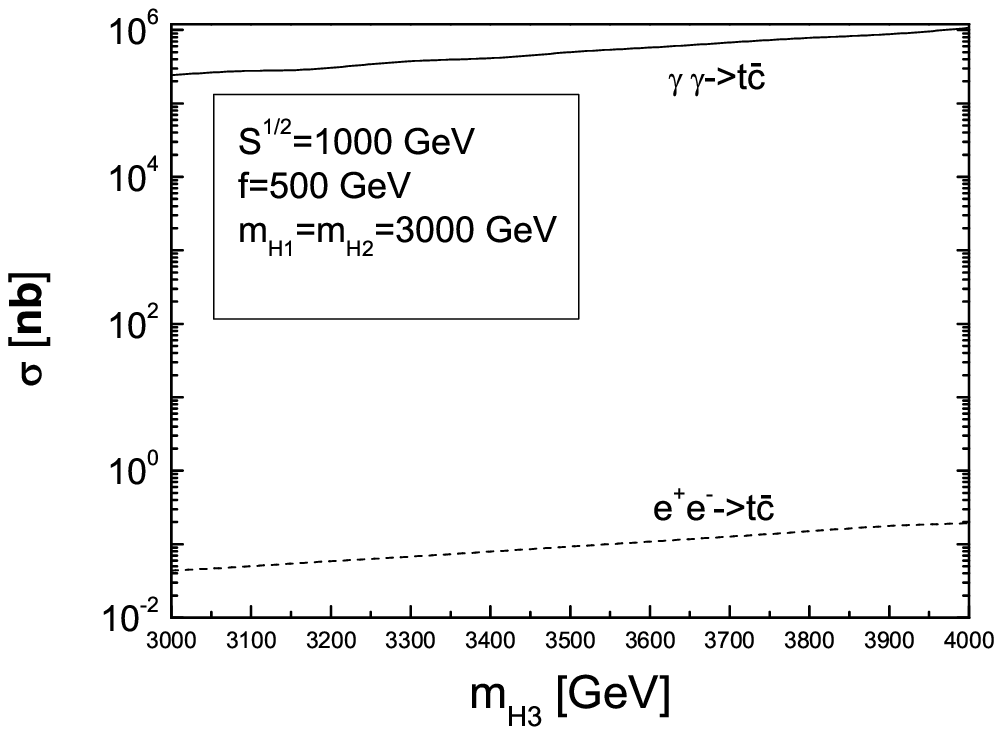}}\\
\caption{\small The cross sections of the processes
$e^+e^-(\gamma\gamma) \rightarrow t\bar{c}$ in the LHT model for
Case II , as a function of $M_{H_3}$.}
\end{figure}

\newpage

\begin{figure}[h]
\scalebox{0.7}{\epsfig{file=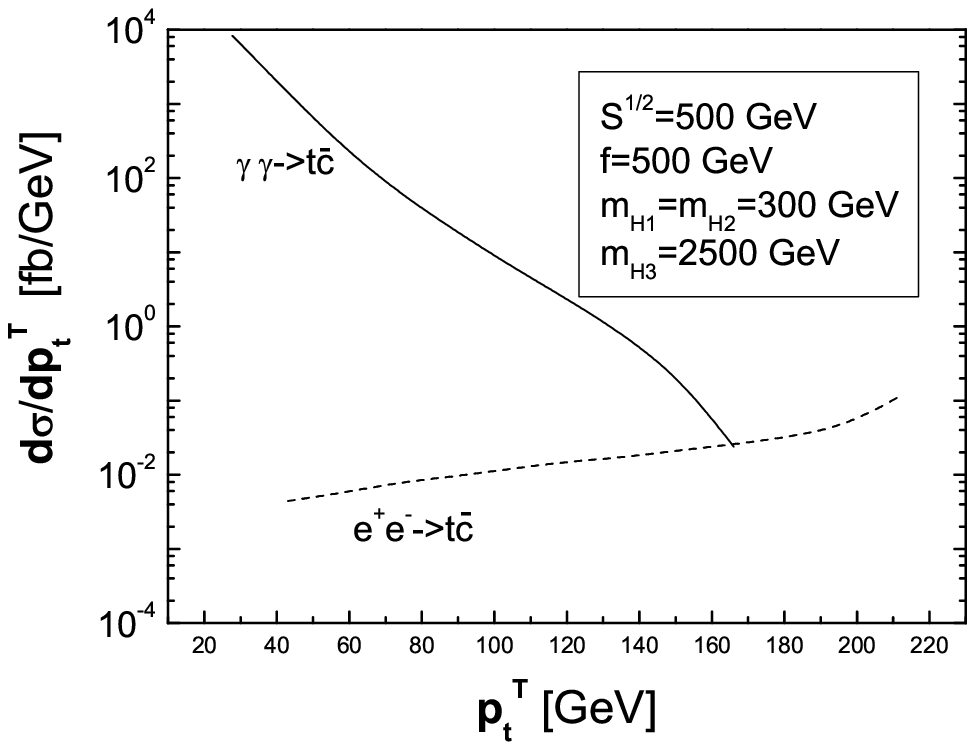}}
\scalebox{0.7}{\epsfig{file=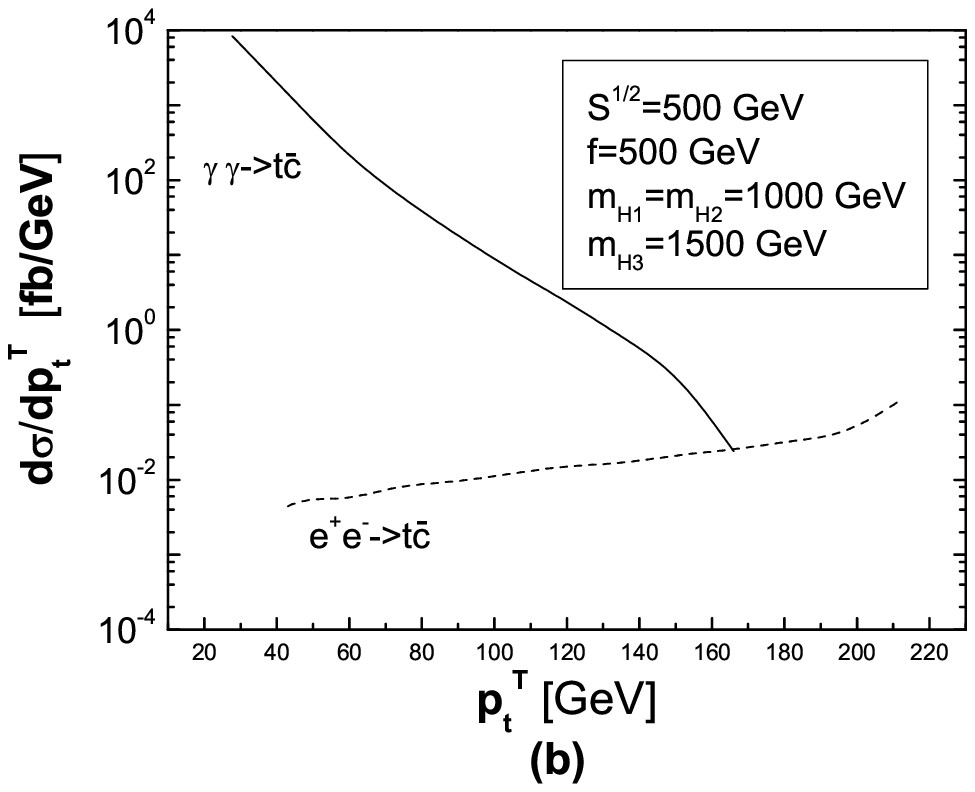}}\\
\caption{\small The transverse momentum distributions of the top
quark for the processes $e^+e^-(\gamma\gamma)\rightarrow t\bar{c}$
in the LHT model. The left diagram is for Case I and the right
diagram is for Case II.}
\end{figure}


\begin{thebibliography}{99}

\bibitem{little Higgs} N. Arkani-Hamed, A. G. Cohen and H. Georgi,  {\it Phys.
Lett.} B{\bf 513}, 232(2001).

\bibitem{LH}
N. Arkani-Hamed, A. G. Cohen, E. Katz, A. E. Nelson, {\it JHEP}
{\bf 0207}, 034(2002).

\bibitem{constraints}
J. L. Hewett, F. J. Petriello, and T. G. Rizzo, {\it JHEP} {\bf
0310}, 062(2003); C. Csaki, J. Hubisz, G. D. Kribs, P. Meade, and
J. Terning, {\it Phys.Rev.} D{\bf 67}, 115002(2003).

\bibitem{LHT}I. Low, {\it JHEP}, {\bf 0410}, 067(2004); H. C. Cheng and I. Low, {\it JHEP}, {\bf 0408}, 061(2004);
J. Hubisz and P.Meade, {\it Phys.Rev.} D{\bf 71}, 035016(2005); J.
Hubisz, S. J. Lee and G. Paz, {\it JHEP}, {\bf 0606}, 041(2006).

\bibitem{scale} J. Hubisz, P. Meade, A. Noble, and M. Perelstein, {\it JHEP}
{\bf 0601}, 136(2006).

\bibitem{He} H. J. He, C. P. Yuan, {\it Phys.Rev.Lett.} {\bf 83},
28(1999); G. Burdman, {\it Phys.Rev.Lett.} {\bf 83}, 2888(1999).

\bibitem{Rare top decay}For FCNC top-quark decay, see, X. L. Wang et.al., {\it Phys.Rev.} D{\bf 50},
5781(1994); C. Yue et.al., {\it Phys.Lett.} B{\bf 508}, 290(2001);
G. R. Lu, F. R. Yin, X. L. Wang, L. D. Wan, {\it Phys.Rev.} D{\bf
68}, 015002(2003); C. S. Li, R. J. Oakes, J. M. Yang, {\it
Phys.Rev.} D{\bf 49}, 293(1994); G. Couture, C. Hamzaoui, H.
Konig, {\it Phys.Rev.} D{\bf 52}, 1713(1995); J. L. Lopez, D. V.
Nanopoulos, R. Rangarajan, {\it Phys.Rev.} D{\bf 56}, 3100(1997);
G. M. de Divitiis, R. Petronzio, L. Silvestrini, {\it Nucl. Phys.}
B{\bf 504}, 45(1997); J. M. Yang, B.-L. Young, X.Zhang,{\it
Phys.Rev.} D{\bf 58}, 055001(1998); J. M. Yang, C. S. Li, {\it
Phys.Rev.} D{\bf 49}, 3412(1994); J. Guasch, J. Sola,{\it Nucl.
Phys.} B{\bf 562}, 3(1999); G. Eilam et al., {\it Phys.Lett.}
 B{\bf 510}, 227(2001); J. J. Liu, C. S. Li, L. L. Yang, L. G. Jin,
{\it Phys.Lett.} B{\bf 599}, 92(2004).

\bibitem{top-charm}For top-charm productions, see,
 J. Cao, Z. Xiong, J. M. Yang, {\it Nucl. Phys.} B{\bf 651},
87(2003); C. S. Li, X. Zhang, S. H. Zhu, {\it Phys. Rev.} D{\bf
60}, 077702(1999); J. J. Liu, C. S. Li, L. L.Yang, L. G. Jin, {\it
Nucl. Phys.} B{\bf 705}, 3(2005); Z. H. Yu, H. Pietschmann, W. G.
Ma, L. Han, Y. Jiang, {\it Eur. Phys.} J. C{\bf 16}, 541(2000); Y.
Jiang, M. L. Zhou, W. G. Ma, L. Han, H. Zhou, M. Han, {\it
Phys.Rev.} D{\bf 57}, 4343(1994); C. Yue, Y. Dai, Q. Xu, G. Liu,
{\it Phys.Lett.} B{\bf 525}, 301(2002); C. Yue, G. Liu, Q. Xu,
{\it Phys.Lett.} B{\bf 509}, 294(2002); C. Yue, G. Lu, J. Cao, J.
Li, G. Liu, {\it Phys.Lett.} B{\bf 496}, 93(2000); J. Cao, Z.
Xiong, J. M. Yang, {\it Phys.Rev.} D{\bf 67}, 071701(2003).

\bibitem{wang}X. L. Wang, Y. L. Yang, B. Z. Li, C. X. Yue, J. Y. Zhang, {\it Phys.Rev.} D{\bf 66},
075009(2002); X. L. Wang, B. Z. Li, Y. L. Yang, {\it Phys.Rev.}
D{\bf 68}, 115003(2003); W. N. Xu, X. L. Wang, Z. J. Xiao, {\it
Eur. Phys. J.} C{\bf 51} 891(2007).

\bibitem{D-LHT} J. Hubisz, S. J. Lee,  and G. Paz, {\it JHEP} {\bf 0606}, 041(2006);
M. Mlanke, A. J. Buras, A. Poschenrieder, C. Tarantino, S. Uhlig
and A. Weiler, {\it JHEP} {\bf 0612}, 003(2006); M. Mlanke, A. J.
Buras, A. Poschenrieder, S. Recksiegel, C. Tarantino, S. Uhlig and
A. Weiler, {\it JHEP} {\bf 0611}, 062(2006).

\bibitem{tcv-LHT}H. S. Hou, hep-ph/0703067

\bibitem{distribution}
G. Jikia, {\it Nucl. Phys.} B{\bf 374}, 83(1992); O. J. P. Eboli,
et~al., {\it Phys. Rev.} D{\bf 47}, 1889(1993); K. M. Cheung, {\em
ibid.} {\bf 47}, 3750 (1993).

\bibitem {HZ}
K. Hagiwara and D. Zeppenfeld, {\it Nucl. Phys}. B{\bf 313},
560(1989); V. Barger, T. Han and D. Zeppenfeld, {\it Phys. Rev.}
D{\bf 41}, 2782(1990).

\bibitem{parameters}Particle Data Group, W. -M.~Yao  {\it et al.}, {\it J.Phys.} G{\bf
33} 1(2006).

 \end{thebibliography}
\end{document}